\documentclass[3p,times,aa,twocolumn]{elsarticle}

\usepackage{ecrc}

\volume{00}

\firstpage{1}

\journalname{Acta Astronautica}

\runauth{R.A. Bamford et al.}

\jid{aa}

\usepackage{amssymb}

\usepackage[figuresright]{rotating}

\usepackage{amsmath,amssymb}
\usepackage{graphicx}
\usepackage{dcolumn}
\usepackage{natbib, textcase,bm}
\usepackage{multicol} 
\usepackage{caption}
\usepackage{subcaption}
\usepackage{url} 

\numberwithin{equation}{section}

\begin{document}

\begin{frontmatter} 


\ead{Ruth.Bamford@stfc.ac.uk}
\ead[url]{www.minimagnetospheres.org}
\cortext[cor1]{Ruth.Bamford@stfc.ac.uk}


\title{An exploration of the effectiveness of artificial mini-magnetospheres as a potential Solar Storm shelter for long term human space missions}


\author[label1]{R.A. Bamford\corref{cor1}}
        \author[label1]{B. Kellett}
        \author[label1]{J. Bradford}
        \author[label2]{T.N. Todd}
        \author[label3]{M. G. Benton, Sr.} 
        \author[label2]{R. Stafford-Allen}
        \author[label4]{E.P. Alves}
        \author[label4]{L. Silva}
        \author[label1]{C. Collingwood}
        \author[label5]{I.A. Crawford} 
        \author[label6,label1]{R. Bingham}
        \address[label1]{RAL Space, STFC, Rutherford Appleton Laboratory, Harwell Oxford, Didcot, OX11 0QX, U.K.}
        \address[label2]{Culham Centre for Fusion Energy, Culham Science Centre, Abingdon, Oxfordshire, OX14 3DB, U.K.}
        \address[label3]{The Boeing Company, El Segundo, CA 90009-2919, USA.} 
        \address[label4]{GoLP/Instituto de Plasmas e Fus$\tilde{a}$o Nuclear, Instituto Superior T\text{\'e}cnico, 1049-001 Lisboa Portugal.}
         \address[label5]{Dept of Earth and Planetary Sciences, Birkbeck College, London}
        \address[label6]{University of Strathclyde, Glasgow, Scotland, UK.}
       

\begin{abstract}

In this paper we explore the effectiveness of an artificial mini-magnetosphere as a potential radiation shelter for long term human space missions. Our study includes the differences that the plasma environment makes to the efficiency of the shielding from the high energy charged particle component of solar and cosmic rays, which radically alters the power requirements. The incoming electrostatic charges are shielded by fields supported by the self captured environmental plasma of the solar wind, potentially augmented with additional density. The artificial magnetic field  generated on board acts as the means of confinement and control. Evidence for similar behaviour  of electromagnetic fields and ionised particles in interplanetary space can be gained by the example of the enhanced shielding effectiveness of naturally occurring "mini-magnetospheres"  on the moon. The shielding effect of surface magnetic fields of the order of $\sim$100s nanoTesla is sufficient to provide effective shielding from solar proton bombardment that culminate in visible discolouration of the lunar regolith known as "lunar swirls". Supporting evidence comes from theory, laboratory experiments and computer simulations that have been obtained on this topic. The result of this work is, hopefully, to provide the tools for a more realistic estimation of the resources versus effectiveness and risk that spacecraft engineers need to work with in designing radiation protection for long-duration human space missions. 

\end{abstract}

\end{frontmatter}  


\hspace{0.3in}
\section{Introduction}
The World’s space agencies are actively drawing up plans for human space missions beyond Low Earth Orbit \cite{ISECG2013}, and scientific benefits resulting from the human exploration of the Moon, Mars and asteroids may be considerable \cite{Spudis1992, Crawford2004, Ehrenfreund2012}. 

\hspace{12mm}

However the risk posed by the radiation in space is one of the major obstacles when considering long term human space exploration \cite{Townsend2005, Lockwood2007, NAP2008}, which means that careful consideration must be given to radiation protection.

The US National Research Council Committee on the Evaluation of Radiation Shielding for Space Exploration\cite{NAP2008} recently stated: 

{\it "Materials used as shielding serve no purpose except to provide their atomic and nuclear constituents as targets to interact with the incident radiation projectiles, and so either remove them from the radiation stream to which individuals are exposed or change the particles' characteristics--energy, charge, and mass -- in ways that reduce their damaging effects."}

This paper outlines one possible way to achieve this, by radically reducing the numbers of particles reaching the spacecraft. The technology concerns the use of `Active' or electromagnetic shielding -- far from a new idea (for a reviews see \cite{Parker2006,Adams2005}) -- but one that has, up until now, been analysed with some crucial factors missing. Specifically the plasma environment of interplanetary space.


So, presented here are the results of asking three questions:
\begin{enumerate}
       \item What difference does the fact that the environment of interplanetary space contains a low density ($\sim10$ per $cm^{-3}$) plasma of positive and negative charges, make to how a potential artificial electromagnetic radiation shield would work on a manned spacecraft?
       \item How differently does a plasma behave at the small scales of a spacecraft compared to, say, a  magnetosphere barrier of a planet? 
       \item How does this change the task of balancing the cost and benefits of countermeasures for the engineers designing an interplanetary or long duration manned mission?
\end{enumerate}

Initiatives such as {\it Earth‐-Moon-‐Mars Radiation Environment Module (EMMREM)}\cite{EMMREM} aim to provide frameworks to overcome the mission safety challenges from Solar Proton/Particle Events (SPEs). But accurate prediction is only of any use if the means to protect the craft and crew actually exist.  

\begin{figure}
\center
        \includegraphics[width=0.45\textwidth]{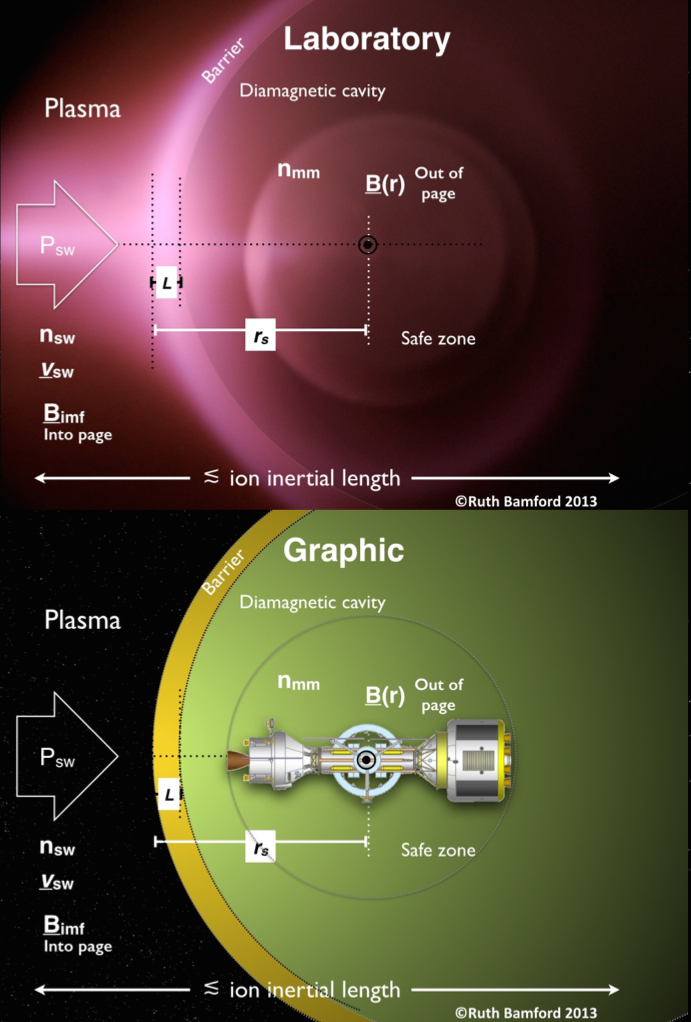}
        \caption{A magnetically held plasma barrier creating an artificial mini-magnetosphere in the laboratory\cite{Bamford2008} (upper panel) and conceptually around a spacecraft (lower panel). Above: The supersonic hydrogen plasma (pink glow) from the Solar Wind Tunnel is coming in from the left hand side and encountering a magnetic field (which inside the protective casing visible in the photograph). The self-captured plasma forms a thin sheath barrier that redirects the incoming hazard. A cavity in the density is created within, confirmed by probe measurements \cite{Bamford2008}. This photograph is taken from above looking through the sheath onto the south pole. The graphic below shows the various zones whose characteristics are discussed within the paper and their relationship to a conceptual spacecraft\cite{Benton2011}.  Importantly in the laboratory experiment the overall dimensions are of the order, or less than, the ion skin depth $c/\omega_{pi}$ which will be essential to be of practical size. The width $L$ of the current sheath is approximately the electron skin depth $L \sim c/\omega_{pe}$.  The pressure balance between incoming and defensive forces occurs at distance ${\bf r_s}$ from the source of magnetic field. Together these parameters define the effectiveness of the active shield. (Conceptual spacecraft design \copyright Mark Benton, Sr.)} 
\label{FIG:mini_mag_ship}
\end{figure}

In this paper we discuss the principles and optimisations specifically of miniature magnetospheres. The upper panel in Figure~\ref{FIG:mini_mag_ship} shows a photograph of a mini-magnetosphere formed in the laboratory \cite{Bamford2008} from the principles outlined in this paper. The application of these principles to the space environment has been shown by comparison between in-situ spacecraft observations of the naturally occurring Lunar mini-magnetospheres\cite{Bamford2012}. Below in Figure~\ref{FIG:mini_mag_ship} is an illustration showing a mini-magnetosphere around a conceptual manned interplanetary spacecraft. 

\subsection{Mini-magnetospheres and plasmas}

In space the charged particles (protons, electrons and other trace ions) mostly originate from the sun and a {\em magnetosphere} is a particular type of `diamagnetic cavity' formed within the {\em plasma} of the solar wind. 

A plasma is a state of matter where the diffuse conglomeration of approximately equal numbers of positive and negative charges is sufficiently hot, that they do not recombine significantly to become neutral particles. Rather the charges remain in a dynamical state of quasi-neutrality interacting, and self-organising, in a fashion dependant upon on the interaction of internal and external electromagnetic forces. It is these attributes that are to be exploited here as a means to protect vulnerable manned spacecraft/bases.

In interplanetary space high energy component of the solar particles is what forms the `hazard' itself because of the high penetrating capability of energetic ions in particular. These are the Solar Cosmic Rays (SCR). There are also a smaller \% (about 6 orders of magnitude less) of super energetic particles at GeV energies that have been accelerated by exotic events like super-novas.n These form the Galactic Cosmic Ray (GCR) component. Both high fluxes of SCR during storms and the long term exposure to GCR are a concern for astronaut health \cite{Cucinotta2006}.

Space plasmas are very diffuse indeed with about 10 particles occupying the volume of the end of the average human thumb, and are considered ultra high vacuum by terrestrial standards. The mean-free-path between physical collisions between the particles is far longer than the system (in solar wind the mean-free-path is about 1 A.U. (Astronomical Unit). This means the particles `collide' through their electrostatic charges and collective movements (such as currents) that are guided by, or result in, magnetic or electric fields. Because of the large dimensions of space, even a very low density is important. The electrostatic forces between two charges are $10^{39}$ times more intense than their gravitational attraction \cite{Alfven1979}. Because a plasma is a rapidly responding (because of the free moving charges), conducting medium it creates a magnetic field in opposition to an externally applied magnetic field making it diamagnetic and can result in local cavities. Diamagnetic cavities are a general phenomenon in plasmas not just in space plasmas and can be formed with or without magnetic fields \cite{Tidman1971}.

Magnetospheres are more generally associated with planetary magnetic fields, such as the Earth's, interacting with the solar wind plasma \cite{Ratcliffe1972}. Miniature magnetospheres are fully formed magnetospheres, with collisionless shocks and diamagnetic cavities, but the whole structure is very much smaller, of the order of 10s to 1000s km across. Mini-magnetospheres have been observed associated with the anomalous patches of surface magnetic field on the Moon\cite{Lin1998}, Mars\cite{Halekas2008} and Mercury\cite{Anderson2011}, and also with asteroids like Gaspra and Ida \cite{Kivelson1993}. It has also been demonstrated that mini-magnetospheres can form without magnetic fields, such as from natural comets \cite{Coates1990} and artificial comets such as AMPTE \cite{Bryant1985}. In these cases the term ``magneto" can still be used because the currents induced in the sheath region include magnetic fields. Mini-magnetospheres are  determined by the plasma physics of the very small scale which in general has been neglected in the analysis of the electromagnetic deflection as a means of spacecraft protection. The entire structures are smaller than the bending radius of an energetic ion about the magnetic field in a vacuum, so this is not a convential 'magnetic shield'.

Presented here is a `block diagram' of the characteristics and parameters needed to implement a mini-magnetosphere deflector shield for a manned space craft. The real physics of the interaction is immensely complicated and largely non-deterministic analytically due to non-linearities. So what is here are `rules of thumb' as guide only. A full detailed analysis requires the use of complex plasma physics and simulations codes best conducted on specific case due to the resources needed.

\section{The hazard}\label{SECTION:Hazard}

At the radius of the Earth’s orbit the level of ultra-violet radiation from the Sun is sufficiently high that photo ionisation results in very little matter in free space remaining unionised. The medium of space is therefore a plasma albeit of very low density. Solar eruptions consist of electromagnetic waves but also protons and electrons with a small percentage of higher mass ions. 

\begin{figure}
 \includegraphics[width=0.45\textwidth]{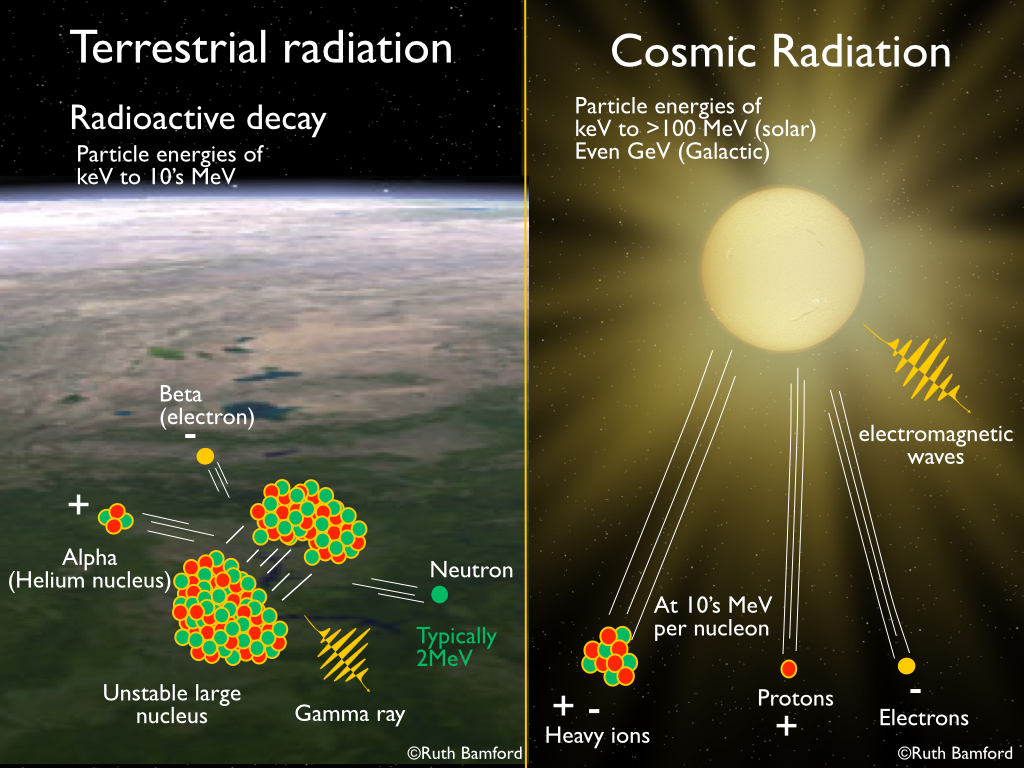}\caption{A different type of ``radiation" in space. Radiation hazard on Earth is generally related to radioactive decay of heavy elements like uranium and electromagnetic waves like gamma and x--rays (left). The radiation in space also has a broadband electromagnetic component but also has an additional form of radiation not seen on Earth except in particle accelerators and as cosmic rays. The forces in space stars, supernovae etc accelerate abundant light elements like hydrogen to MeV to 100's GeV energies. At these energies the particles are predominantly the nuclei of atoms and electrons separately constituting a high energy plasma.}\label{FIG:TypesRadiation}
\end{figure}

\begin{figure}
\center
 \includegraphics[width=0.41\textwidth]{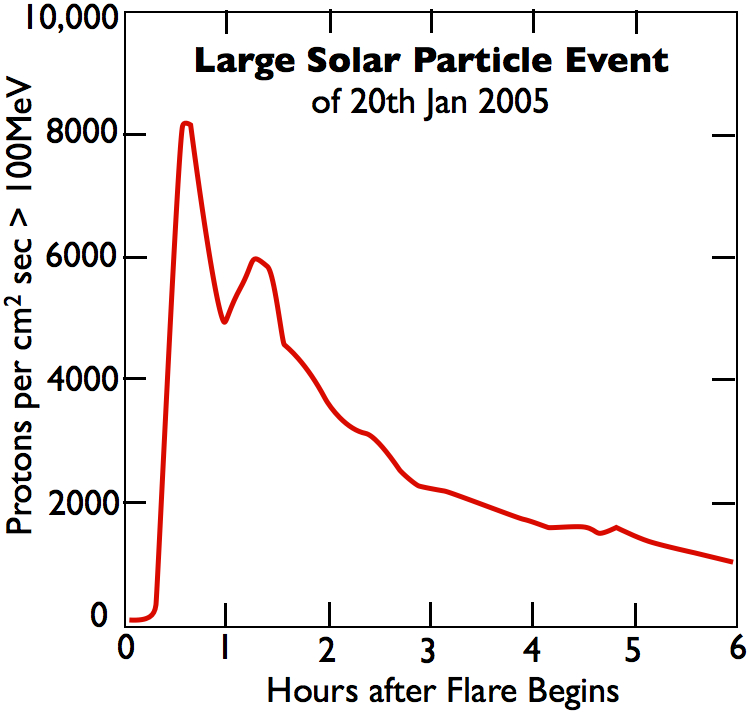}
 \includegraphics[width=0.45\textwidth]{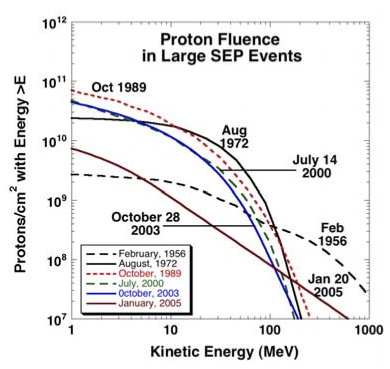}
        \caption{(a) The time evolution of proton fluence for a large storm. High-energy $> 0.5$GeV protons arrived at Earth, peaking within 5 minutes followed 10 minutes later by the peak $>100$MeV fluence, $\sim \times 10,000$ the background level. Over the next 12 hours directional, lower energy particles arrive still at elevated densities above quiet times. (b) The energy fluence spectra of some of the largest SEP events of the last 50 years \cite{Mewaldt2005}. Under normal conditions the numbers of particles with energies $>$10MeV is negligible.
But increases of 10 to 1000 times during a SEP are typical and can rise to as much as $10^6$ for more extreme events \cite{Mewaldt2005}.}\label{FIG:Pin}
\end{figure} 

The radiation encountered in space (see Figure~\ref{FIG:TypesRadiation}) is a composite of a small percentage of extremely high energy galactic particles, a higher density but, much lower energy continuous outflow of particles from the sun (the solar wind), interspersed with intermittent, high density eruptions of very energetic particles originating from a variety of violent events on the sun. Events on or near the sun that result in shockwaves can accelerate ions and electrons to extremely high energies \cite{Cane1988a, McClements1997}. An examples showing the temporal and energy spectra of large Solar Energetic Particle (SEP) events is shown in Figure~\ref{FIG:Pin} \cite{Mewaldt2005}.

One of the more recent large Solar Energetic Particle (SEP) events illustrates the magnitude of the problem \cite{Mewaldt2005}. The temporal plot of particle flux from \cite{Mewaldt2005} is shown in Figure~\ref{FIG:Pin}(a). The x-ray flare on the Sun provided only a few minutes warning before high-energy $> 0.5$GeV protons arrived at Earth, peaking within 5 minutes.  Approximately 10 minutes later, the peak in the $>100$MeV protons arrived. At its peak the particle flux rate is $\sim \times 10,000$ the background level. A second peak occurred about 90 minutes later. Over the next 12 hours directional, lower energy particles arrive still at elevated densities above quiet times. 

For a spacecraft in interplanetary space, this results in intense bursts of radiation of deeply penetrating particles capable of passing through the hull to the crew inside. The result is a significant increase for dose-rates above 0.05 Gy/h \cite{Wu2008}.

The variable shape of the energy spectrum for each SPE is an extremely important factor for the total exposure calculation and not just the total fluence~\ref{Kim2010}. For instance, protons with energies $>30MeV$ can pass through space suits, above $~70-100MeV$ then hull walls of $5 --10 g cm^{-2}$ aluminium can be penetrated with the added consequences of secondariness. The energy spectrum of some of the largest events (based on fluence of particles) of the last 50 years is shown in Figure~\ref{FIG:Pin}(b) \cite{Mewaldt2005}. 

The vulnerability of different organs and systems (such as Blood Forming Organs or nervous system) varies considerably\cite{Cucinotta2001,Cucinotta2010,Borak2014}. Thus it becomes difficult to quantify the potential mission disruption caused solar events based purely on predicted size of event.

Current estimates~\cite{Kim2010} suggest that there is  $\sim$20\% chance of exceeding the current NASA 30-d limit for a future SPE with $\Phi_{30}= 2\times10^9 protons cm^{-2}$ on an in interplanetary journey.The probability of multiple events increases with mission period. 

Protection against extremely large solar energetic particle events (SPE) that sporadically occur with very little warning, is a mission critical issue for long term, interplanetary manned missions \cite{Townsend2005, Lockwood2007}.

\subsection{Particle description of the incoming Pressure}

The characteristics of the instantaneous plasma (quasi-neutral collection of approximately equal numbers of positive and negative charges), particle distributions impacting the spacecraft define how the plasma shield will function at any one instant. 

The pressure from the environmental plasma $P_{in}$ can consist of more than one component. Considering a thermal part and a bulk flow ram pressures, and pressure from the magnetic field in the solar wind: 
\begin{equation}
          P_{in} = P_{th,sw} + P_{ram,sw} +P_{B_{IMF}}+P_{ram,++}
\end{equation}\label{EQU:Pin}
The component terms being $P_{th} = n_{th}kT_{th}$, $P_{ram} = n_{sw}m_{sw}{\bf v_{sw}}^2$ and $ P_{B_{IMF}} = |{\bf B_{IMF}}^2|/2\mu_o$. Here $n_{sw}$ represents the density of particles flowing at at velocity ${\bf v_{sw}}$ and ${\bf B_{IMF}}$ is the interplanetary magnetic field (IMF). The final term is the ram pressure from the high energy particles $P_{ram,++}$ which has been differentiated from the main distribution for this analysis. As can be seen from Figure~\ref{FIG:Pin}(b) the density of particles at the high energy tail can be a significant fraction of the bulk density but in general can be considered as negligible fraction of the pressure.

As will be seen in the following sections the ever present, though variable, background `solar wind' plasma is what is used to initially {\em create the barrier}. As will be shown in later sections it can then be augmented artificially if necessary to increase the deflection of the hazardous high energy part of the particle spectrum.

\section{Mini-magnetospheres}

\subsection{Pressure Balance}
The principle of `Active Shielding' requires electromagnetic forces to balance the incoming pressure. (Many authors have reviewed the general principles of Active Shielding, for example \cite{Parker2006}).

We start with a generic expression of the required pressure balance:
\begin{equation}
          P_{mm} = P_{th, mm} + P_{ram, mm} +P_{B_{mm}}
\end{equation}\label{EQU:Pmm}

Here the subscripts represent the pressures from within the mini-magnetosphere. In practice the ram component, although it exists due to the motion of the spacecraft, is just a frame of reference issue and insignificant. In many cases the thermal outward pressure is also insignificant. So to first order a magnetosphere pressure balance provides:
\begin{equation}
         P_{B_{mm}} \approx P_{ram, sw}. 
\end{equation}

As will be shown later there is an electric field generated by the formation of a mini-magnetosphere, so this term will remain. If we create an artificial mini-magnetosphere on board the spacecraft we can define the initial $P_{B_{mm}}$.

\section{Creating an artificial mini-magnetosphere}

An on-board Mini-Mag system would most likely be comprised of a superconducting coil \cite{French1968}. 

In a non-conductive medium, the magnetic field intensity of a dipole magnetic field diminishes rapidly with range. Higher-order structures, such as quadrapoles and octopoles, have fields which fall off much more rapidly with radius from the coils that create them.

\begin{figure}
    \centering
    \includegraphics[width=0.45\textwidth]{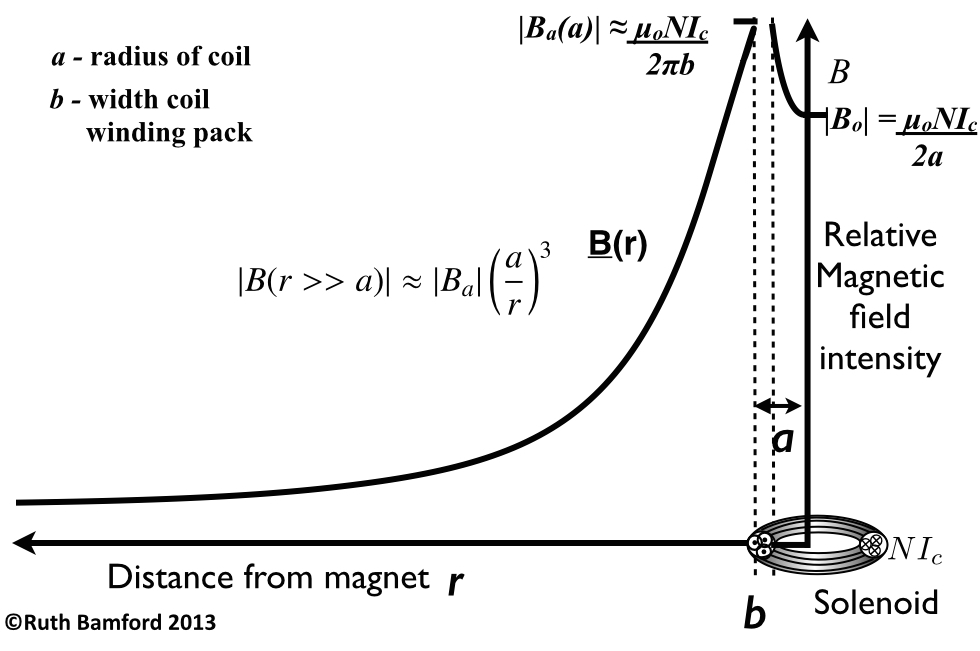}%
    \caption{How the vacuum magnetic field intensity varies with range from the spacecraft based on solenoid characteristics. The magnetic field of a flat round coil of major radius $=a$, with $N$ turns of current carrying windings of $I_c$ the current in each turn results in a total amp-turns $I$ of $I=NI_c$, provides a magnetic field intensity at the centre of the coil of $B_o$. The highest magnetic field is $B_o(b)$ and is at the surface of the winding pack where $b$ is the minor radius winding pack. As can be seen from the figure the magnetic field intensity in the center can be less than the outer edge of the winding pack. The wider the radius the more this is the case, and the greater the range field beyond the spacecraft. A central dipole magnetic or multiple magnets (multipole field) the magnetic field intensity will drop off much more rapidly than a wide diameter loop.}
\end{figure}\label{FIG:B_vac}

\begin{figure}
        \centering
        \includegraphics[width=0.45\textwidth]{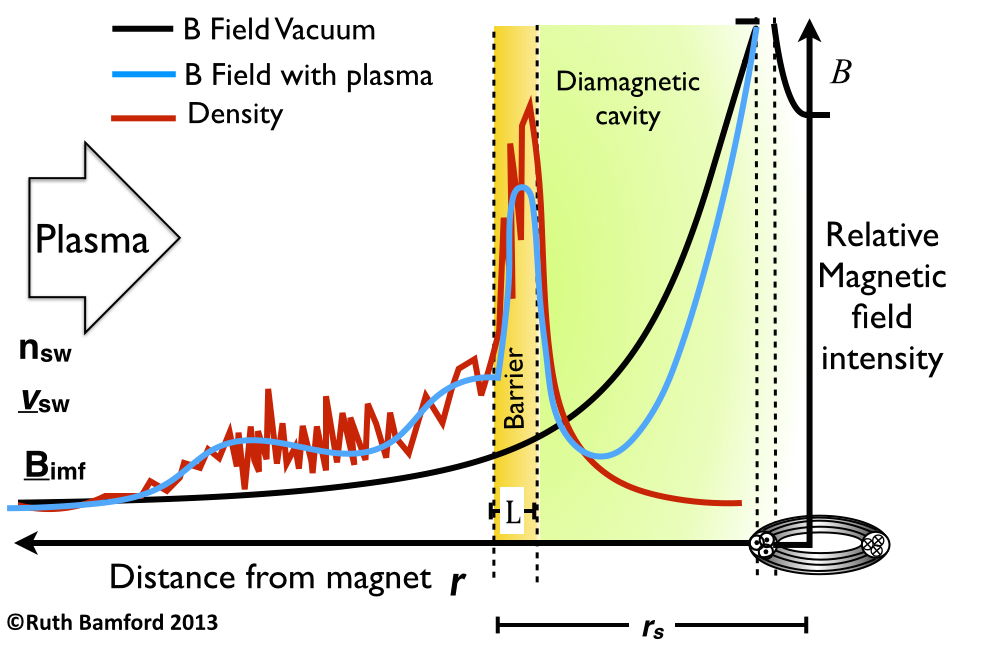}
        \caption{A plot illustrating the difference the plasma environment makes. Sketch of the vacuum magnetic fields when the  plasma environment is included. The foot region is caused by ion reflection and is of the order of the ion inertial length whereas the current or barrier layer width, L, is associated with the electron inertial length. Ideally $r_s \geqslant L$. Because the interaction is a collisionless-shock, the initial pile-up of density and magnetic field is accompanied by both turbulence and a reduction in the velocity of the ions and changes in temperature of both the ions and electrons. Inside the barrier region is the cavity where the population of energetic particles is reduced. To optimise logistics this need be only as wide as to afford sufficient protection as required.}
\end{figure}\label{FIG:B_profile}

Figure~\ref{FIG:B_vac} shows the magnetic field at a distance (far field) where $r~\gg~a$ (in any direction) is $|B_{vac}(r)| \approx |B_o (a/r)|^3$ but only when no plasma is present. Or in terms of the current in the coil:
\begin{equation}
      B(r) \approx \left ( \frac{\mu_oI}{2a} \right ) \left ( \frac{a}{r} \right )^3
\end{equation}\label{EQU:B_r}
Here $I$ is the total loop--current of the solenoid $I=N I_c$, where $N$ is the total number of turns carrying current $I_c$ at radius of $a$.

The presence of the plasma changes the profile. This can be seen illustrated in Figure~\ref{FIG:B_profile}. 

The prohibitively high power estimates of a magnetic shield are based on the vacuum profile (the near field plot generation and profile with distance is shown in Figure~\ref{FIG:B_vac}). The vacuum field power estimates do not allow for the alteration in the profile and additional force illustrated in Figure~\ref{FIG:B_profile}. 

The effect of the plasma environment is not just to extend the range of the magnetic field intensity. The effect of the magnetic `pile-up' comes with cross field currents in a narrow barrier region (or shell in 3D) some distance from the spacecraft. These currents and accompanying electric fields alter how the incoming plasma is deflected. The efficiency of the shielding is therefore found to be much greater than the initial vacuum calculation would have predicted. Evidence that this is the case will be shown in section \ref{SEC:Evidence}.

Quantifying the level of the enhancement and the effectiveness at deflecting higher energy particles is non-trivial. In the following section we shall determine some estimates that can be used to determine the value of an artificial mini-magnetosphere shield for astronaut protection.

\section{Characterising the effectiveness of an artificial mini-magnetosphere}

Figure~\ref{FIG:mini_mag_ship} shows a two dimensional sketch of the morphology of a mini-magnetosphere about a spacecraft. The size of the mini-magnetosphere is dependant on two parameters. 

Firstly, $r_s$ the `stagnation' or `stand-off' distance of a magnetopause is where the pressure of the incoming plasma, $P_{in}$, is balanced by the combined pressure of the mini-magnetosphere. 

The second parameter is $L$ the width of the magnetopause boundary. Clearly to be within the safety of a mini-magnetosphere diamagnetic cavity one must be further away than the thickness of the boundary. 

In kinetic studies of mini-magnetospheres we find that $L\approx$ electron skin depth.

\subsection{Calculating stand-off distance,$r_s$}

To first order equation\eqref{EQU:Pin}, the magnetosphere pressure balance can be approximately taken to be balance by the vacuum magnetic field: $P_{in} \approx P_B$ this occurs at a distance $r_s$ from the source of the magnetic field. In planetary magnetospheres $r_s$ would be the Chapman–Ferraro distance. The calculation is the same but the source of the magnetic field from an artificial source can be included to better relate the effectiveness to power requirements on board.

For dipole magnetic field produced by a solenoid, such as that  shown in Figure~\ref{FIG:B_vac}, the magnetic field at the center of the solenoid $B_o=\mu_oNI_c/2a$, where $a$ is the radius of the solenoid loop containing total loop turns of $I = NI_c$ ($N$ number of turns, $I_c$ current per turn). This provides:
\begin{equation}
     r_s^6 \sim \frac{\mu_o (NI_c)^2}{8P_{in}}a^4
\end{equation}\label{EQU:r_s} 
Here $P_{in}$ is obtained from equation~\ref{EQU:Pin}.
This same calculation for the Earth's magnetosphere, leads to a consistent under-estimation of the true stand-off distance indicating the importance of the other terms in Equations~\eqref{EQU:Pin} and \eqref{EQU:Pmm}. 

Interestingly equation~\eqref{EQU:r_s} reveals that the largest achievable stand-off distance  is achieved with the largest possible coil radius. This is intuitively reasonable because the long-range field strength goes like $B_oa^3$, so a small change in the radius of the coil has a large effect. 

\subsection{Calculating barrier width, $L$}

The plasma physics of the interaction \cite{Tidman1971, Woods1987} tells us that the width of the magnetopause boundary $L$ is of the order of the electron skin depth $\lambda_{e}$.
\begin{equation}
            L \approx \lambda_{e} = \frac{c}{\omega_{pe}}
\end{equation}\label{EQU:skin_depth}
Here $\omega_{pe}$ is the electron plasma frequency, $\omega_{pe}= \left ( \frac{n_e e^2}{\epsilon_om_e} \right)^{1/2}$, $c$ is the speed of light. 

The classical skin depth is a rapid decay of electromagnetic fields with depth inside a conductor caused by eddy currents in the conductor. High frequencies and high conductivity shorten the skin depth as does an increase in the number of current carriers (plasma density).

The same is true here with some differences, for instance the conditions of a collisionless shock of the mini-magnetosphere means the attenuation profile is closer to a linear approximation profile than the $1/e$ attenuation in metals. 

\subsection{The Normalised Linear attenuation factor, $\alpha$}
 
We can now introduce a geometric parameter $\alpha$ as {\it quasi} linear attenuation factor. This is to provide an indication of relative effectiveness. 

The number of skin depths required for complete ambient plasma exclusion is not generally known but values of 4-6 have been calculated up to relativistic energies \cite{Phelps1973}. This is similar to the exponential\footnote{The plasma physics here provides a more linear rather than exponential drop off.} form of the electromagnetic skin depth in metals. However given the level of other approximations being made in these formulations we shall take a normalised parameter where the number of required skin depths is taken to be $=1$.
\begin{equation}
      \alpha := \frac{r_s}{L}
\end{equation}\label{EQU:alpha}
The plasma for this calculation can be a combination of the incoming plasma density and any additional density added from the spacecraft to enhance the shield effectiveness.

The assumed value of $r_s = L$ is good, $r_s >L$ would be better if multiple skin depths $r_s \sim 4-6\times L$, and $r_s < L $ is less than optimum.

\subsection{The origin of the electric field}\label{SEC:Er}

The expressions above provide an estimation of how the bulk pressures balance. For a practical shield to reduce the penetrating high energy component we need to determine the value of the electric field within the barrier - and how it can effect the higher energy exclusion. Figure~\ref{FIG:DeflectionGeometry} is a close up view of Figure~\ref{FIG:mini_mag_ship} with the force vectors overlaid. The particle trajectory of a representative high energy ion velocity ${\bf v_{++}}$ and mass $M$ is also shown.

The forces on the charged particle are determined by the Lorentz force: 
\begin{equation}
    {\bf F} = q({\bf E} + {\bf v} \times {\bf B}).
\end{equation}\label{EQU:Lorentz}
Unlike in a vacuum, the presence of the plasma means that the ${\bf E}$ cannot be neglected and is related to ${\bf B}$.

The electric field component comes from the formation of a currents that are induced to exclude the interplanetary magnetic field and create the cavity. 

The physics of collisionless shocks provides us with an expression for the instantaneous electric potential component, $\phi$, responsible for slowing and deflecting the ions such that:
\begin{equation}
	 \phi(r)\approx - \frac{\kappa}{n_{mm}}\frac{|\Delta B_{mm}|^2}{\Delta r}
\end{equation}\label{EQU:phi}
Here $\kappa$ is a constant $\kappa =1/(2\mu_0e)$. If $|\Delta B_{mm}|\sim|B_{mm}|$ is the intensity of the magnetic field orthogonal to ${\bf r}$ and ${\bf v}$ at distance $(r_s -L)$, then $\Delta r \sim L$. 

Equation~\eqref{EQU:phi} shows that the potential is related to the {\em gradient} in the magnetic field intensity, or the {\em pondermotive force}. This is a much more effective and short--range force than calculations of the magnetic bending alone would suggest. Because the density within the mini-magnetosphere, $n_{mm}$ is within both $\kappa$ and $L$ (from equation~\eqref{EQU:skin_depth} ), the magnitude of $\phi \propto \sqrtsign{n_{mm}}$. This offers a means to boost the effectiveness of the deflector shield by adding additional density. This shall be discussed in Section~\ref{Sec:Boosting}.

\subsection{High Energy particle deflection}\label{Sec:HighEnergy}

The electric field that is created is responsible for changing the energy and trajectory of the energetic particles. 

Although the electric field values from equation~\ref{EQU:phi}, even with augmented plasma density, are not going to be sufficient to stop $>100MeV$ or $GeV$ ion, this is not required. The particles need only be {\em refracted} sufficiently away from the central safe--zone. Much like defending against a charging rugby-footballer, rather than stand in his way to protect the goal line, a better policy is to {\em deflect} the player sideways using a small amount of force so he is pushed into touch and out of the field of play.

\begin{figure}
        \includegraphics[width=0.45\textwidth]{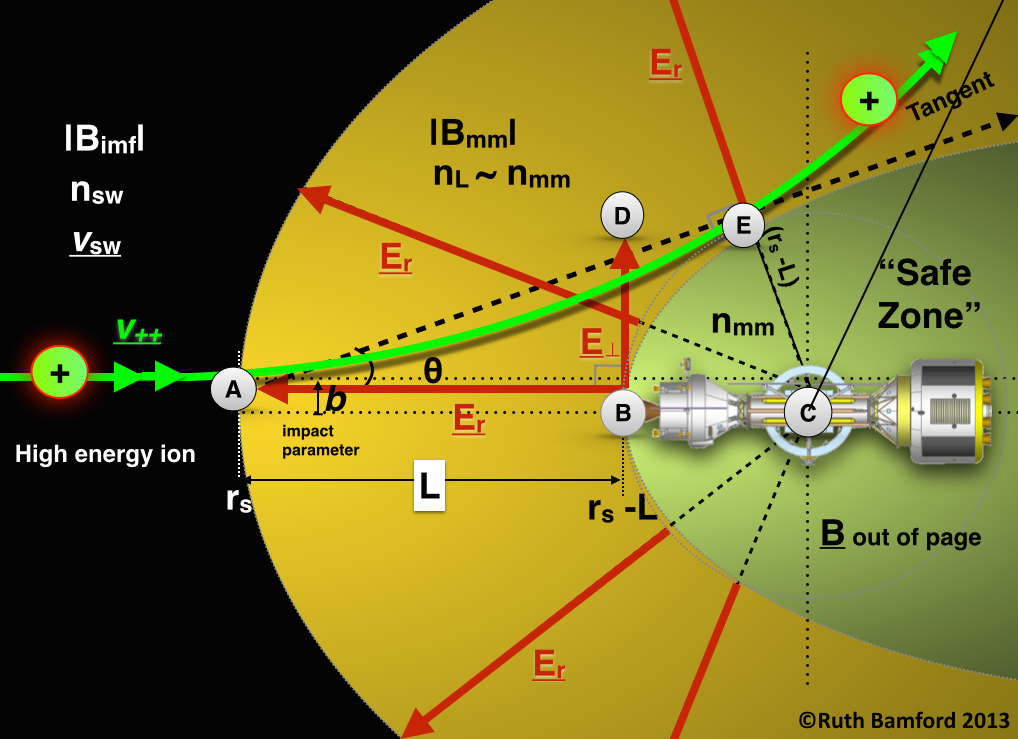}
      \caption{The deflection of a high energy ion (green) by the electric field ($E_r$) (red) created by the low energy plasma captured and retained by the magnetic field from the spacecraft. Augmenting the natural density, by releasing readily ionised gas from the craft, can enable protection of $>>100MeV/amu$ ions.}\label{FIG:DeflectionGeometry}
\end{figure}

The geometry of this for our case is illustrated in Figure~\ref{FIG:DeflectionGeometry}.

In Cartesian coordinates, the deflection component of the electric field, $E_\perp$, required to just miss the spacecraft, is acquired across the whole barrier width, $L$, in the one plane is $|E_\perp |= |E_r| \tan \theta$ where $\theta$ is the angle of the charged particle to the radial or the scattering angle.

The needed deflection velocity $v_{\perp}$ becomes:
\begin{equation}
    v^2_{\perp} \approx \frac{\kappa}{n_{mm}M} \frac{|\Delta B^2_{mm}|}{L}
\end{equation}\label{EQU:deflection_velocity}

As has already been mentioned in Section~\ref{SEC:Er}, in 3D the physics is such that the electric field will always point outwards away from the spacecraft. This results in a 3D safe zone effective against both directional and omni-directional threats. Thus we must determine the effectiveness of the high energy scattering process.

Because the electric field is formed self-consistently by the plasma itself, and the high energy particles are scattered by a lower electric field, the problem of generating a secondary population of ions accelerated towards the spacecraft by the deflector shield itself, does not occur. 

The mini-magnetosphere barrier interaction with high energy particles is far from simple.The incoming high energy particle not only sees the electric field set up by the interaction of the solar wind and the spacecraft field, it also experiences the usual convective electric field as seen by a charged particle moving relative to a magnetic field. This convective field ($E_{\perp}$) is perpendicular to the magnetic field. This results in the particle being deflected by a series of fields in a complex manner \cite{Woods1987}. 

Quantifying the shield performance for specific spectra of high energy particles (like an SEP) requires a full 3D recreation using a computer simulation, or an experiment either in space or in the laboratory. 

\subsubsection{Computer simulation of high energy scattering}

\begin{figure}
        \includegraphics[width=0.45\textwidth]
{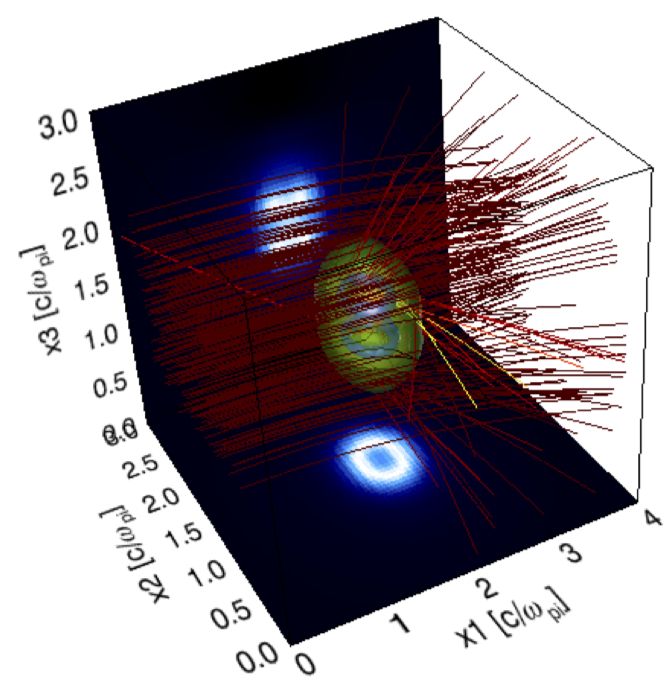}
\caption{A simulation of particle tracks (red) scattered from a thin electrostatic ``shell'' (green) surrounding a magnetic dipole (center with B field intensity projected onto faces of the cube). The particles are not being deflected by the magnetic field rather by the electric field created by the interaction of the background plasma (not shown for clarity) and the magnetic field. The energy of the red `protons' is $100,000$ times that of the background plasma. Simulation in dimensionless units.
}\label{FIG:Scatter}
\end{figure}

Figure~\ref{FIG:Scatter} shows a simulation of high energy scattering from a dipole magnetic field (center of box). The 3000 `SEP' incoming ions are $100,000$ the energy of the thermal background plasma ($\sim5eV$) contained within the box (the 'solar wind'). These simulations showed that 100\% of particles were excluded from the ``safe zone", whilst 95\% were excluded for particles $\times \sim 10^6$.

This indicates a narrow electric field is responsible for the deflection and not the gradual bending due to a magnetic field.

\subsection{Boosting the shield effectiveness: Mass loading}\label{Sec:Boosting}

The very biggest storms could be mitigated against by adding additional plasma density around the spacecraft. Similar to creating an artificial cometary halo cloud. 

Increasing the density within the mini--magnetosphere reduces the thickness of the skin depth (equ.~\eqref{EQU:alpha}). This could be done either to reduce the power required from the space craft to achieve the same deflection efficiency, or boost the shield effectiveness during the severe parts of a SEP or CME event. 

Practically this could be done by releasing easily ionised material from the spacecraft. EUV ionisation, charge exchange, and collisional ionisation lead to the generation of ions and electrons which are incorporated in the mini-magnetosphere barrier. The mass loading leads to enhancement in the currents. 

For manned spacecraft utilising Nuclear Thermal Propulsion (NTP) or Nuclear Electric Propulsion (NEP) as primary propulsion this would already be achieved by the release of propellant from the thrusters; typical propellants for these systems being hydrogen and other volatile propellants, and for NEP systems, also inert gases such as Argon and Xenon \cite{Guyen2011,long2012,Goebel2008}. If more localised injection of plasma is required toward the shield region, ion or plasma sources, as already used for spacecraft propulsion \cite{Goebel2008}, could be used to provide more directed ion or plasma beams from multiple locations on the spacecraft.
 
During the transit to Mars it might be necessary to use the augmented storm shield upon $0 \sim2$ occasions \cite{Kim2010}. Increasing $n_{mm}$ by $\times 10^4$ would provide an increase the potential $\phi\sim\times 100$. This could be achieved by approximately 1~mole of Xe (in the volume of $\sim \frac{4}{3}\pi r_s^3$). Given the atomic mass of Xe is $\sim 131$ this would mean 131g of Xe would be needed per occasion of use. 

Exactly how much Xe would be needed on a mission would depend upon the frequency of use. Allowing for approximately 3 SEP events to encompass the spacecraft in 18 month period this would require less than half a kilo of Xe.

It would also be required to keep the enhancement sustained for 2-6 hours. How much resources will be required would then depend upon the rate of plasma loss from the mini--magnetosphere. This is discussed in the next section.

\subsection{Retaining the Shield}\label{Sec:Retain} 

To function as a shield it is required that enough density within the cavity barrier is retained for long enough, to ensure the cavity  will not get over whelmed by an intense storm for the hours that the peak fluence may last(Figure~\ref{FIG:Pin} (a)). The plasma parameter, $\beta$, defined as the ratio of the plasma pressure to the magnetic pressure, does not provide a useful guide in this instance because the profiles of plasma density and temperature are varying on spacial scales below the ion gyration radius. Furthermore the parameter $\beta$ does not allow for electric fields which we know are fundamental to the mini-magnetosphere barrier. 

Since an analytical approach is not available as a guide, we can take an observational example from comets \cite{Galeev1985,Coates1990}, and in particular the AMPTE artificial comet\cite{Bryant1985}. The data recorded by the spacecraft monitoring the active magnetospheric particle tracer explorers (AMPTE) mission, provide us with a lower limit of retainment in the absence of a magnetic field. A $\sim$ 1Kg mass of Barium (amu=137) exhibits an ionisation time of $\sim$20 minutes for a volume of 100km \cite{Luhr1986}. With the cometary case the particle pick-up means the confinement structure is essentially open ended and the matter is lost rapidly. The addition of a magnetic field would undoubtably extend the plasma retention (as is the case with magnetically confined plasma fusion experiments such as JET \cite{JET}) but by precisely how much, particularly on the scale size of a mini-magnetosphere, could only be determined experimentally in space.

\section{Estimating the requirements of the on board hardware}

Having outlined the principles behind the mini-magnetosphere shield operation, and assembled some performance parameters, we can now compute some figures of merit. 
 
A conceptual deep space vehicle for human exploration described in \cite{Benton2011} included a mini-magnetosphere radiation shield. The purpose was to present a candidate vehicle concept to accomplish a potential manned near-Earth object asteroid exploration mission. The power (including cyroplant), physical dimensions and magnetic field intensities and density augmentation capabilities used here will be those presented in \cite{Benton2011, Benton2012}.

The maximum feasible coil radius, $a$ of 3.0m (set by the launch rocket cowling I/D), $I_c = 700A, N= 8000$ resulting from equation~\eqref{EQU:B_r} in $NI_c = 5.6$MAt. This produces a peak magnetic field, $B_o$, of $\sim 6.4$T. Inserting all the practical parameters provides $r_s =0.86/P_{in}^{1/6}$ in units of  km and nPa.

The total mini-magnetosphere power demand limit of $16kW $, and a 5kW for the cryoplant and the control system. The total mass was $ = 1.5 \times 10^3kg$. 

\begin{figure}
        \includegraphics[width=0.45\textwidth]{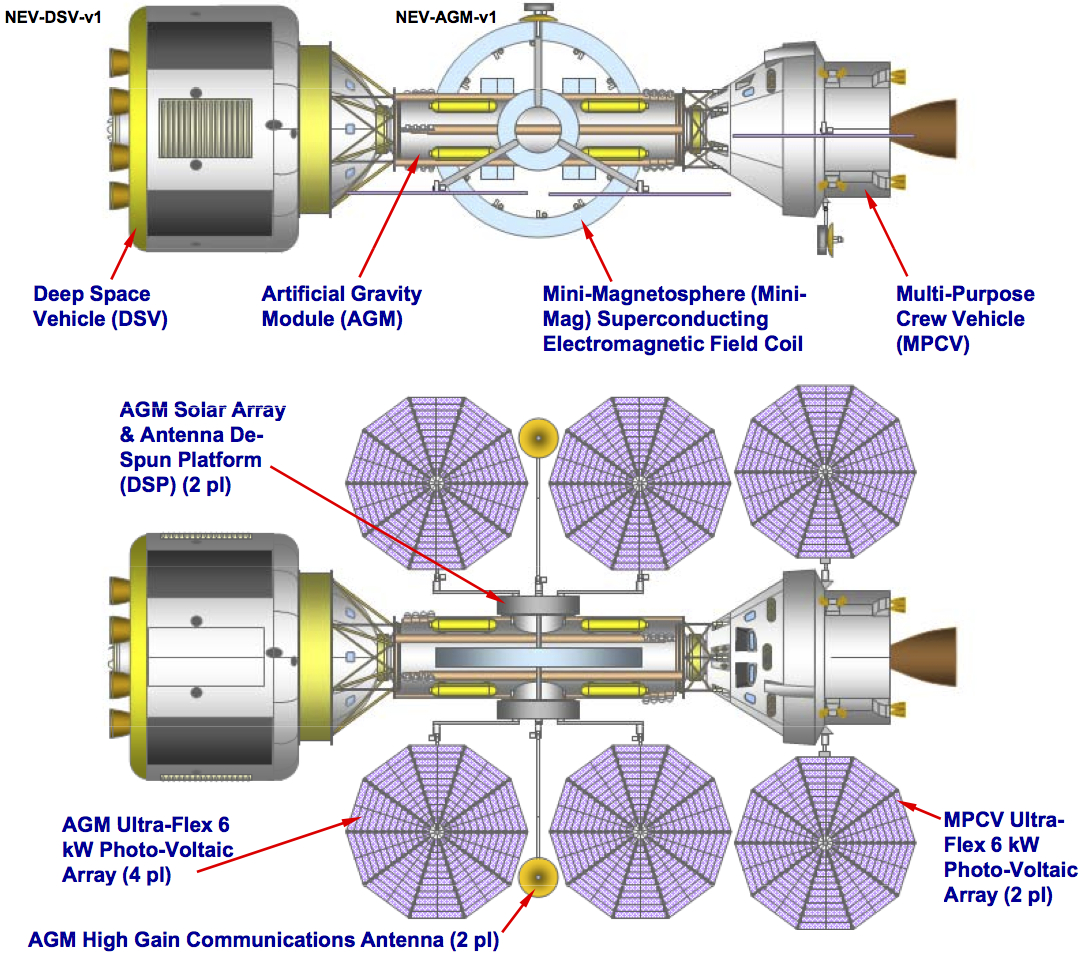}
        \caption{A conceptual design for a manned interplanetary vehicle was presented by Benton \cite{Benton2013} using current established technology and incorporated a mini-mag system. } \label{FIG:Benton}
\end{figure}

\section{Evidence for the processes from other fields}\label{SEC:Evidence}

The experimental and observational evidence for the formation of mini-magnetospheres  has been established from laboratory using Solar Wind Plasma Tunnels \cite{Bamford2008, Gargate2008} and spacecraft observations of natural mini-magnetospheres on the Moon \cite{Lin1998, Bamford2012}.  

\begin{figure}
                \includegraphics[width=0.4\textwidth]{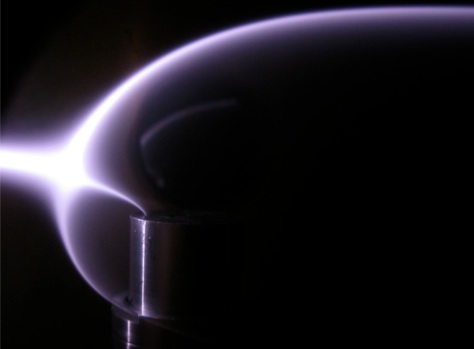}
 \caption{A photograph of a mini-magnetosphere diamagnetic cavity formed in a laboratory Solar Wind Tunnel\cite{Bamford2008}.The `light' areas show where the plasma is present. The beam comes in from the left hand side and gets redirected into a thin layer around the target.The width of the layer $L$ and the stand-off distance $r_s$ agree very well with those expected by equ.\ref{EQU:r_s} and \ref{EQU:skin_depth}}\label{FIG:LabBubble}
 \end{figure}

A photograph of a laboratory sized mini-magnetosphere is shown in Figure~\ref{FIG:LabBubble}~\cite{Bamford2008}. A vacuum or an MHD (magnetohydrodynamic) description of the laboratory experiment would have predicted that the plasma stream would not be deflected and hit the magnet.

\section{Summary}

The equations provided above, can only provide ``ball-park'' values as the complexity of the interaction is highly variable with multiple parameters inter-dependant in time and orientation. This is typical description of a non-linear system. We know that mini-magnetospheres work because of the example on the Moon \cite{Lin1998,Bamford2012}. We know that the same principles used here occur for both natural and artificial comets \cite{Galeev1985,Bingham1991}.   

The addition of extra, cold, plasma, such as Xenon gas which can be easily ionised by the uv-radiation from the Sun, can potentially greatly increase the effectiveness of the shield.

The concept of having a plasma around the spacecraft may at first sound familiar to those looking at Active shield systems. \cite{French1968} amongst others have proposed various 'Plasma Shield' schemes of using flowing currents in plasmas around the spacecraft as means to extend the magnetic field or source of electrons to counter the incoming protons. The difficulty with these schemes has been the omission of what the environmental plasma would do by way of short circuiting, reactive screening or `blowing away' of the plasma of the mini-magnetosphere by the solar wind. The scheme suggested here does not try to control the plasma too much but confine it {\it enough} to allow it's own nature to work for us. 

Regardless of whether some of the details contained in this paper can be improved or adapted, this paper has aimed to emphasise the importance of including the plasma environment when considering any means of Active or electromagnetic shielding to protect spacecraft from ionising radiation. 

This paper has also aimed to demonstrate the importance of using the appropriate plasma physics dominant at the ``human'' rather than ``celestial'' scale size.  

We have tried to provide ball-park expressions to estimate a realistic prediction of effectiveness that are credible and not underplay the complexity and research to be done. 

The analysis shown here is for a modest powered mini-magnetosphere system that may function as a permanent means to increase the operating time in interplanetary space for crew and systems, much like the Earth's magnetosphere does. Such a shield could also be bolstered to deal with extreme storms, for which it maybe the only means of providing an effective storm shelter.

\section{Conclusions}

Proposals for electromagnetic shields  generally come with  amazing predictions of effectiveness. And yet no such system has even been tried as a prototype in space. The reason for this is that the claims of effectiveness appear unbelievable - quite rightly. 

What has been presented in this paper, is an indication of the true complexity involved in active shielding. Simple back-of-the-envelope calculations for the particle deflection in a vacuum are vastly over-simplistic. Inevitably the role of the plasma environment has either been overlooked completely, or analysed on the wrong scale size. 

Much has yet to be determined quantifiably before the full engineering standard of precision is available.It may be that an active shield system is not practical until on-board power systems are comparable to those envisioned in science fiction, but the concept should not be dismissed based on an incorrect analysis. 

An active deflector shield system would never be a replacement for passive shielding nor biological advances. But it can offer options, particularly for EVAs, extending the longevity of hardware, preventing secondary activation of the ships hull and systems and the only theoretical means to deflect even GeV particles. 

The evidence that mini-magnetosphere do work on the bulk plasma in space  comes from magnetic anomalies  on the moon \cite{Bamford2012, Bamford2014}, around asteroids \cite{Kivelson1993}, comets both natural \cite{Coates1990} and artificial\cite{Luhr1986}. This combined with laboratory experiments \cite{Muggli2001} and simulations have suggested that the high energy distribution can be sufficiently effected to enable optimism, because even the ability to predict the occurrence of storms is of little use if there is nowhere to hide.

\section{Acknowledgements}
The authors would like to thank Science and Technology Facilities Research Council's Center for Fundamental Physics. 


\bibliographystyle{elsarticle-num}
%


\end{document}